\documentclass[aps,reprint,showpacs,floatfix,amsmath,amssymb,pre]{revtex4-1}

\usepackage{pdfpages}

\usepackage{lmodern}						

\begin{document}

\title{Collective dynamics in two populations of noisy oscillators \\ with asymmetric interactions}
\author{Bernard Sonnenschein$^{1}$}
\author{Thomas K. DM. Peron$^{2,3}$}
\author{Francisco A. Rodrigues$^{4}$}
\author{J\"urgen Kurths$^{1,3}$}
\author{Lutz Schimansky-Geier$^{1}$}
\affiliation{$^{1}$Department of Physics, Humboldt-Universit\"at zu Berlin, Newtonstrasse 15, 12489 Berlin, Germany}
\affiliation{$^{2}$Instituto de F\'isica de S\~ao Carlos, Universidade de S\~ao Paulo, CP 369, 13560-970 S\~ao Carlos, S\~ao Paulo, Brazil}
\affiliation{$^{3}$Potsdam Institute for Climate Impact Research (PIK), 14473 Potsdam, Germany}
\affiliation{$^{4}$Instituto de Ci\^encias Matem\'aticas e de Computa\c{c}\~ao, Universidade de S\~ao Paulo, CP 668, 13560-970 S\~ao Carlos, S\~ao Paulo, Brazil}

\begin{abstract}
We study two intertwined globally coupled networks of noisy Kuramoto phase oscillators that have the same natural frequency, but differ in their perception of the mean field and their contribution to it. Such a give-and-take mechanism is given by asymmetric in- and out-coupling strengths which can be both positive and negative. We uncover in this minimal network of networks intriguing patterns of discordance, where the ensemble splits into two clusters separated by a constant phase lag. If it differs from $\pi$, then traveling wave solutions emerge. We observe a second route to traveling waves via traditional one-cluster states. Bistability is found between the various collective states. Analytical results and bifurcation diagrams are derived with a reduced system.
\end{abstract}
\maketitle

\section{Introduction}
From neurons in the brain to cells in the heart, from electrons in superconductors to planets in the universe, 
collective oscillations are ubiquitous \cite{strogatz2003sync,*PikRosKu03}. Clearly, the underlying mutual synchronization 
critically relies on the presence of interactions. If individuals are allowed to interact non-uniformly, the collective behavior remains 
particularly elusive. Beyond incoherence and synchronization, what other collective states can emerge, and under which conditions?
Motivated by Daido's seminal work on ``oscillator glasses'' \cite{daido1992quasientrainment}, Hong and Strogatz tackled this question recently in a series of papers \cite{HoStr12,HoStr11PRL,*HoStr11}. They investigated two mutually globally coupled populations of Kuramoto phase oscillators that differed in their coupling strengths. Two scenarios of mixed attractive and repulsive interactions were distinguished. In the first case some oscillators' phases repel the phases of all the others, while the remaining attract all the other phases \cite{HoStr12}. This situation resembles neural networks with excitatory and inhibitory connections \cite{wilson1972excitatory,*brunel2000dynamics}. In the second case some oscillators tend to align with the mean field, while others oppose it favoring an antiphase alignment \cite{HoStr11PRL,*HoStr11}. This version is analogous to sociophysical models of opinion formation \cite{Miguel2005}. Surprisingly, only the second scenario led to enriched dynamics beyond the traditional order-disorder transition.

In this paper, we unify both coupling scenarios. We unveil the existence of two routes to traveling waves, which in parameter space are surrounded either by states of diametral two-cluster synchronization or by one-cluster partially synchronous states. Differently than in Refs. \cite{HoStr12,HoStr11PRL,*HoStr11} we consider temporal fluctuations acting on the frequencies. Our main contribution is the derivation of bifurcation diagrams for all possible collective states.

\section{Model}
One of the most prominent models describing phenomena of mutual synchronization is due to Kuramoto \cite{Kur84}.
It describes how the phases of coupled oscillators evolve in time, and is applicable to systems of nearly identical, weakly coupled 
limit-cycle oscillators. We consider a stochastic version with twofold disordered coupling strengths:
\begin{equation}
\dot{\phi}_i(t)=\omega_0+\frac{K_i}{N}\sum_{j=1}^{N}G_j\sin\left(\phi_j-\phi_i\right)+\xi_i(t),
\label{model}
\end{equation}
where $i=1,\ldots,N$. Using the notion of give-and-take as a metaphor, oscillator $i$ contributes to the mean field with weight $G_i$ and at the same time incorporates the mean activity with weight $K_i$ into its own dynamics. Accordingly, we call $K_i$ in- and $G_i$ out-coupling strength, respectively. Grouping together oscillators with the same coupling strengths, the number of different pairs $\left(K_i,G_i\right)$ coincides with the number of subpopulations. 

We consider two equally-sized subpopulations denoted by ``$1$" and ``$2$." Hence, the oscillators are distinguished by a pair of coupling strengths, $\left(K_1, G_1\right)$ or $\left(K_2, G_2\right)$, and all of those can be positive or negative. We choose the parametrization
\begin{equation}
K_{1,2}=K_0\pm\frac{\Delta K}{2},\ \ G_{1,2}=G_0\pm\frac{\Delta G}{2}. 
\label{parametrization}
\end{equation}
$K_0$ and $G_0$ are average in- and out-coupling strengths, while $\Delta K$ and $\Delta G$ give corresponding mismatches. If $\left|\Delta K\right|/2>\left|K_0\right|$ or $\left|\Delta G\right|/2>\left|G_0\right|$, then half of the couplings are positive (attractive) and half negative (repulsive). In such cases we speak of mixed interactions.
Note that \eqref{parametrization} leads to point symmetries, because changing $\left(K_0,G_0\right)\rightarrow\left(-K_0,-G_0\right)$ or 
$\left(\Delta K,\Delta G\right)\rightarrow\left(-\Delta K,-\Delta G\right)$ yields the same situations.
In the following, all oscillators have the same constant natural frequency 
$\omega_0$. Therefore, by virtue of the rotational symmetry, we can set $\omega_0=0$ without loss of generality. Time-dependent disorder $\xi_i(t)$ is included as Gaussian white noise, 
$
\langle\xi_i(t)\rangle=0,\ 
\langle\xi_i(t)\xi_j(t')\rangle=2D\delta_{ij}\delta(t-t').
$
The angular brackets denote averages over different realizations of the noise and the single non-negative parameter $D$ denotes the noise intensity. 
The noise terms $\xi_i(t)$ can be regarded as an aggregation of various stochastic processes \cite{AniAstNeVaSchi07}. In Refs. \cite{HoStr12,HoStr11PRL,*HoStr11} it is found for the deterministic case $D=0$ that mixed out-couplings alone do not enable more than partial synchronization, whereas mixed in-couplings yield traveling waves reached through diametrically synchronized states. We intermingle both types of mixing and explore in particular whether traveling waves persist in the presence of noise $D>0$. Without loss of generality all subsequent results are obtained with $D=0.5$, but for illustration we keep $D$ in the derivations.

``Discordant synchronization" is used here as an umbrella term for situations where the ensemble splits into two partially synchronized clusters. This will include traveling waves and $\pi$-states, the latter being the extreme form of discordance with two oscillator populations anti-aligned to each other.

\section{Theory}
We investigate the thermodynamic limit $N\rightarrow\infty$, where 
propagation of molecular chaos \cite{kac1954foundations} allows us to describe each population by a \emph{one-oscillator} probability density 
$\rho_{1,2}(\phi,t)\equiv\rho(\phi,t|K_{1,2},G_{1,2})$. Normalization requires $\int_0^{2\pi}\rho_{1,2}(\phi,t) d\phi'=1\ \forall\ t$. 
For given coupling strengths $K_{1,2}$ and $G_{1,2}$, $\rho_{1,2}(\phi,t) d\phi$ denotes the fraction of oscillators 
with phase between $\phi$ and $\phi+d\phi$ at time $t$. The densities are governed by the nonlinear 
Fokker-Planck equations \cite{Sak88,*StrMir91}:
\begin{equation}
   \frac{\partial\rho_{1,2}}{\partial t}\,=\,D\frac{\partial^2\rho_{1,2}}{\partial\phi^2}-\frac{\partial}{\partial\phi}\left[K_{1,2}R\sin\left(\Theta-\phi\right)\rho_{1,2}\right]\ .
 \label{fpe}
 \end{equation}
The global mean-field amplitude $R(t)$ and phase $\Theta(t)$ follow from a superposition:
 \begin{equation}
  R(t)\mathrm e^{i\Theta(t)}=\frac{1}{2}\left[r_1(t)\ G_1\ \mathrm e^{i\Theta_1(t)} + r_2(t)\ G_2\ \mathrm e^{i\Theta_2(t)}\right],
  \label{order_fpe}
\end{equation}
Note that subpopulations of different sizes can be treated simply by rescaling $G_{1,2}$.
The local mean-field variables obey
\begin{equation}
r_{1,2}(t)\mathrm e^{i\Theta_{1,2}(t)}=\int_{0}^{2\pi}d\phi'\mathrm e^{i\phi'} \rho_{1,2}\left(\phi',t\right).
  \label{localorder}
\end{equation}
The level of synchrony in the two subpopulations is measured separately by $r_{1,2}(t)$, whereas for the global measure we take the classical
Kuramoto order parameter 
$
r(t)\equiv\frac{1}{2}\left|r_1(t)\ \mathrm e^{i\Theta_1(t)} + r_2(t)\ \mathrm e^{i\Theta_2(t)}\right|,
$
which differs from Eq. \eqref{order_fpe} by taking out $G_{1,2}$. The order parameter values lie between zero (incoherence) and $1$ (complete synchronization). The variables $\Theta(t)$ and $\Theta_{1,2}(t)$
stand for the corresponding mean phases. Of special interest is the phase lag $\delta(t)$, i.e. the difference in the mean phases
of the two populations, $\delta(t)=\Theta_1(t)-\Theta_2(t)$.
\begin{figure}[!tpb]
\begin{center}
\includegraphics[width=0.98\linewidth]{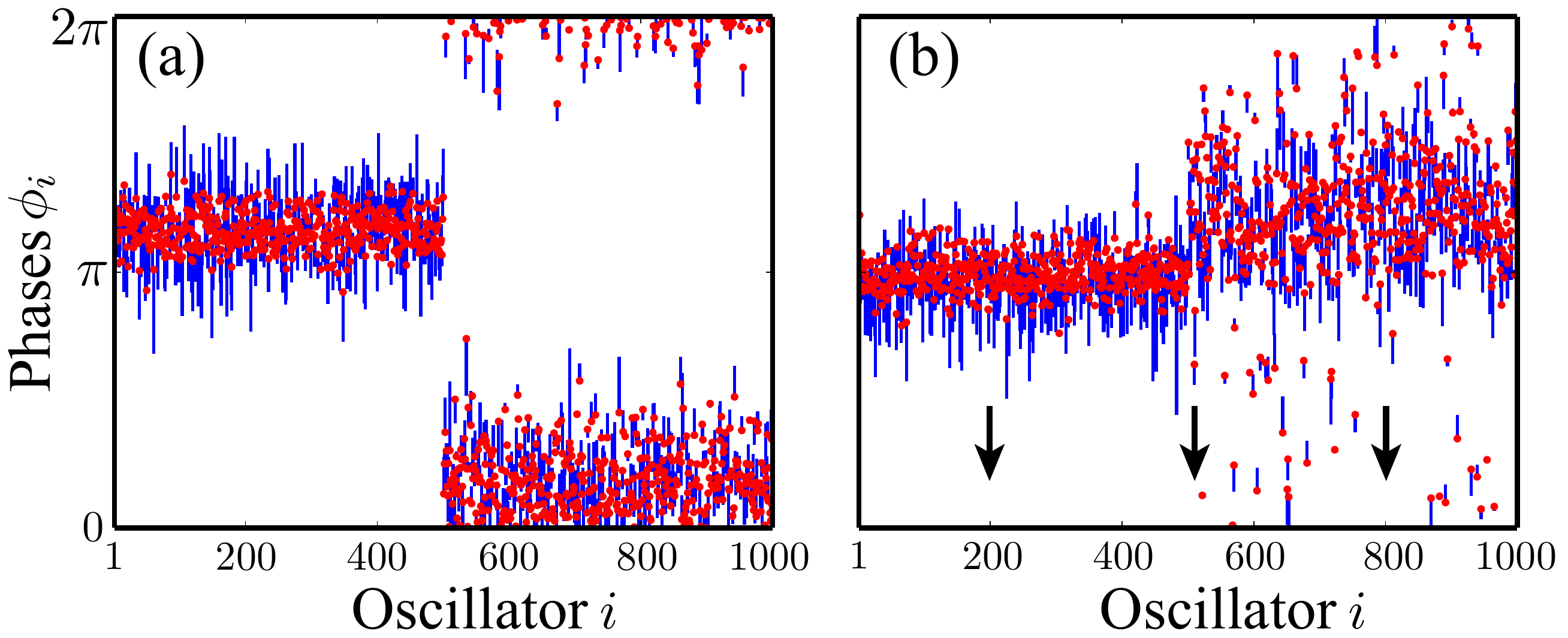}
\end{center}
\caption{(Color online) Snapshots of oscillators' phases. 
(a) Stationary $\pi$-state at $K_0=1$, mean phase difference $\delta=\pi$, no net flux; 
(b) Traveling wave at $K_0=3$, flow downwards with stationary profile, indicated by black arrows;
red dots depict phase values, blue lines of different lengths and directions indicate instantaneous frequencies.
Remaining parameters: $\Delta K=3$, $G_0=2$, $\Delta G=10$, $D=0.5$. 
For visualization the two subpopulations are separated into two halves.}
\label{Fig1}
\end{figure}

In the sequel, variables without a dot or other indicated time dependence refer to the long-time limit.
Let us outline the four qualitatively different self-organized states observed here after some transient dynamics. 
(i) In the incoherent state the whole population of oscillators rotates asynchronously, $r_{1,2}=0$. 
(ii) The classical partially synchronized state has zero phase lag, $r_{1,2}>0,\ \delta=0$. We use ``zero-lag sync'' as a shortcut to denote this state.
(iii) The $\pi$-state describes a partially synchronized state, where the two subpopulations are anti-aligned to each other, $r_{1,2}>0,\ \delta=\pi$. 
(iv) In the traveling wave state the whole population is also partially synchronized, but oscillates with a frequency different from the frequency
of single oscillators. This spontaneous change in rhythm is induced by a phase lag that is neither zero nor $\pi$, $r_{1,2}>0,\ 0<\delta<\pi$.
We calculate the wave speed as 
\begin{equation}
\Omega=\frac{1}{N}\sum_{i=1}^N\langle\dot\phi_i\rangle_t,
\label{wavespeed} 
\end{equation}
where $\langle\ldots\rangle_t$ represents a long-time average \cite{HoStr11PRL,*HoStr11}.

In Fig. \ref{Fig1} example snapshots from simulations of $N=10^5$ oscillators are shown ($N=10^3$ are equidistantly chosen for visualization). 
Figure \ref{Fig1}(a) displays a $\pi$-state with $r_1 = 0.98$, $r_2 = 0.89$, $r = 0.05$, $\left|\delta\right|=\pi$ and $\left|\Omega\right|=0$. 
Figure \ref{Fig1}(b) shows a traveling wave state with $r_1 = 0.98$, $r_2 = 0.75$, $r = 0.80$, $\left|\delta\right|=1.42$ and $\left|\Omega\right|=3.17$. It is equally possible that the wave runs in the other direction, depending on initial phases and realization of the noise. Note that perfect synchrony, $r_{1,2}=1$, cannot be achieved with finite coupling strengths, if an infinitesimal amount of noise is present.

In order to analytically investigate the collective dynamics that are governed by \eqref{fpe}-\eqref{localorder}, we approximate the phase distributions in the two populations by time-dependent Gaussians. 
This well-known method is motivated by numerical observations \cite{KurSchu95,*ZaNeFeSch03,SonnSchi13}.
It has been generalized recently to encompass heterogeneities in couplings \cite{SonnSchi13,SonnZaNeiLSG13,*SoPeRoKuLSG14}. 
Extending those derivations to the present case, we obtain the following three-dimensional system of ODE's:
\begin{equation}
\begin{aligned}
\dot{r}_{1}=&-r_{1}D+\frac{1-r_{1}^4}{4}K_{1}\left[r_{1}G_{1}+r_{2}G_{2}\cos\delta\right],\\
\dot{r}_{2}=&-r_{2}D+\frac{1-r_{2}^4}{4}K_{2}\left[r_{2}G_{2}+r_{1}G_{1}\cos\delta\right],\\
\dot{\delta}=&-\frac{\sin\delta}{4}\left[\left(r_{1}^{-1}+r_{1}^3\right) K_{1} r_{2} G_{2} + \left(r_{2}^{-1}+r_{2}^3\right) K_{2} r_{1} G_{1}\right].
\end{aligned}
\label{rtheta}
\end{equation}
All the four aforementioned collective states are fixed points of \eqref{rtheta} with $\dot{r}_{1,2}=\dot{\delta}=0$. Two types of fixed point solutions have to be distinguished, because there are two possibilities that $\dot{\delta}=0$ holds: 
\begin{eqnarray}
\delta &=& m\pi,\ m\in\mathbb Z, \label{third_cond_1} \\
0 &=& \left(r_{1}^{-1}+r_{1}^3\right) K_{1} r_{2} G_{2} + \left(r_{2}^{-1}+r_{2}^3\right) K_{2} r_{1} G_{1}. \label{third_cond_2}
\end{eqnarray}
\begin{figure}[!tpb]
\begin{center}
\includegraphics[width=0.99\linewidth]{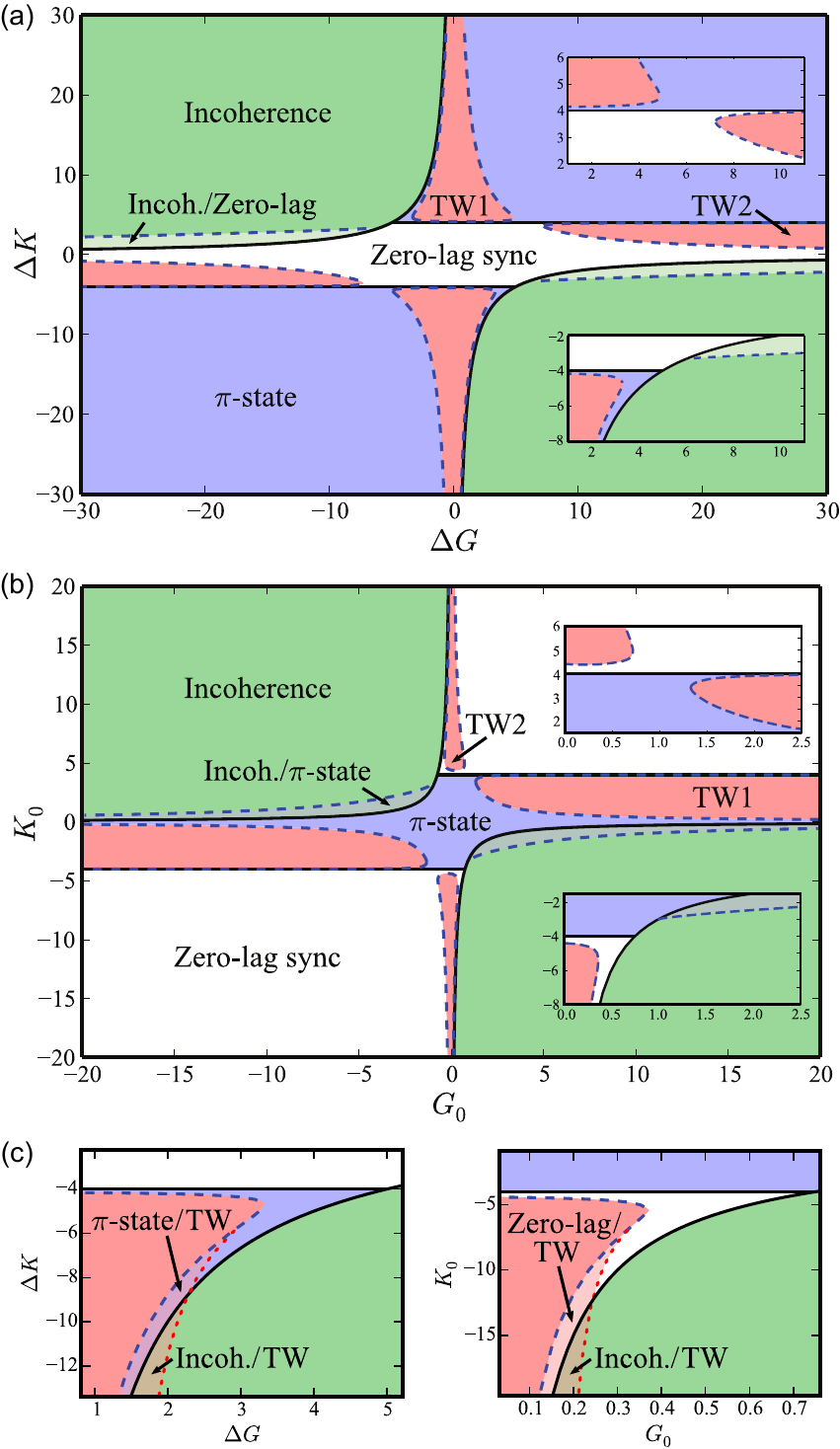}
\end{center}
\caption{(Color online) Bifurcation diagrams for (a) $K_0=2$, $G_0=3$ and (b) $\Delta K=8$, $\Delta G=2$. 
Solid lines from Eqs. \eqref{Kcrit12}, \eqref{Kcrit3}, dashed lines with $\texttt{MATCONT}$ \cite{DhGoKuz03}. ``Zero-lag sync'' denotes partially synchronous states with zero phase lag between the two subpopulations. Parameter regions with lag $\delta=\pi$ labeled as ``$\pi$-states''. 
Traveling waves (TW1 and TW2) have non-zero wave speed $\Omega$. Insets show enlarged areas. (c) Further zooms show coexistence of traveling waves with other collective states. Red dotted lines obtained with $\texttt{MATCONT}$, plotted only here.}
\label{Fig2}
\end{figure}
Equation \eqref{third_cond_1} describes zero-lag and $\pi$-states, whereas Eq. \eqref{third_cond_2} underlies traveling waves. 
Intermediate phase lags $0<\delta<\pi$ cause spontaneous drifts, because according to Eqs. \eqref{rtheta} and \eqref{third_cond_2} the common frequency of the traveling waves obeys
\begin{equation}
\lim_{t \to \infty}\dot\Theta_1=\lim_{t \to \infty}\dot\Theta_2=\sin\delta\ \frac{r_2^{-1}+r_2^3}{4}\ K_2 G_1 r_1.
\label{common_freq}
\end{equation}

Two more equations are obtained from imposing $\dot{r}_{1,2}=0$ in \eqref{rtheta}:
\begin{equation}
\begin{aligned}
  r_{1} =& \frac{r_{2}}{G_{1}\cos\delta}\left[\frac{4D}{\left(1-r_{2}^4\right)K_{2}}-G_{2}\right],\\
	r_{2} =& \frac{r_{1}}{G_{2}\cos\delta}\left[\frac{4D}{\left(1-r_{1}^4\right)K_{1}}-G_{1}\right].
 \label{r2}
\end{aligned}
\end{equation}
With \eqref{third_cond_1}, \eqref{third_cond_2} and \eqref{r2} we have three coupled equations for three unknowns, $r_{1,2}$ and $\delta$.
No stationary solution with $\delta=\pi/2$ can be found, but the singularities $G_{2,1}=0$ and $K_{1,2}=0$ turn out to have a special meaning.
In particular, if one of the in-coupling strengths $K_{1,2}$ vanishes, the corresponding population remains incoherent.
Numerical continuation around this point shows that in order to avoid a negative local order parameter, $r_{1,2}<0$, which is unphysical, the whole population transfers to a $\pi$-state. In our parametrization this first critical condition can be written as
\begin{equation}
\Delta K_{c_{1}}=\pm 2K_0. \label{Kcrit12}
\end{equation}
One can show that in general the incoherent state, $r_{1,2}\equiv 0$, loses linear stability, if the noise intensity falls below a certain value (cf. the Appendix). This happens at
\begin{equation}
\left(\Delta K \Delta G\right)_{c_{2}}=8D-4K_0G_0. \label{Kcrit3}
\end{equation}
\begin{figure*}[!tpb]
\begin{center}
\includegraphics[width=0.95\linewidth]{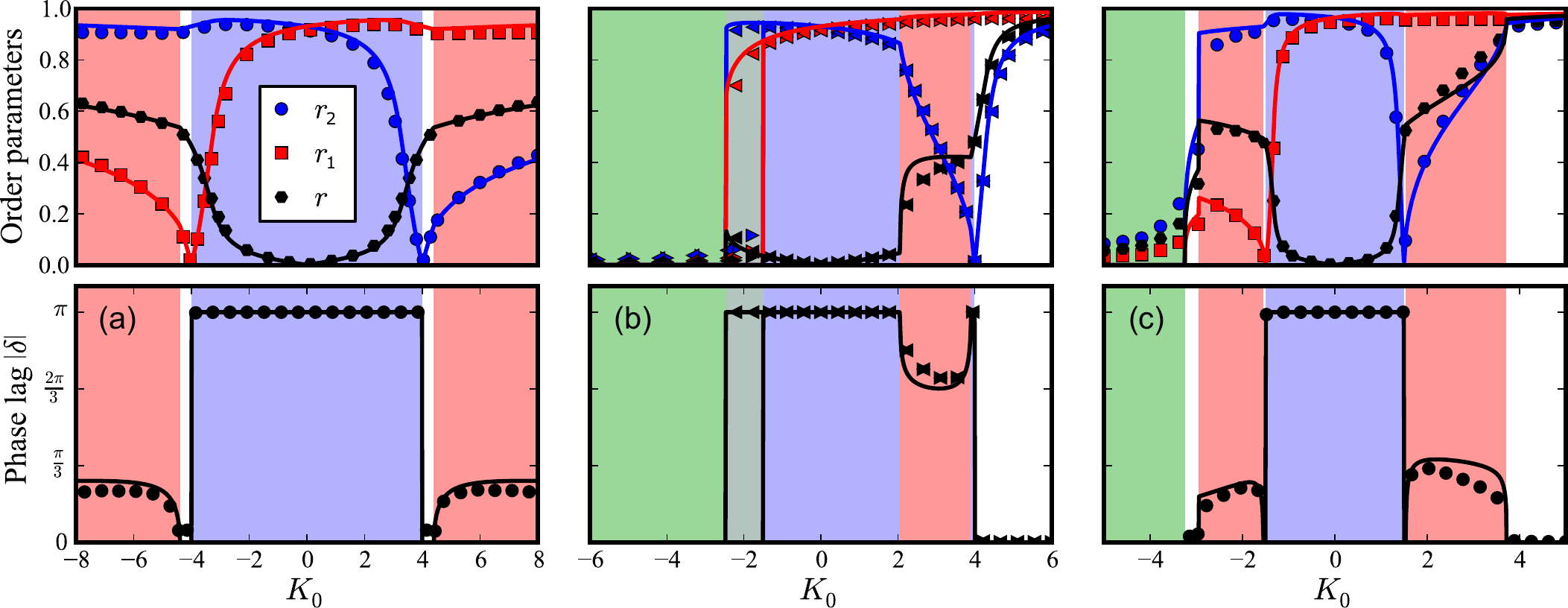}
\end{center}
\caption{(Color online)  Order parameters $r_2, r_1, r$ and  phase lag $\left|\delta\right|$ as a function of average in-coupling strength $K_0$. 
(a) $G_0 = 0$, (b) $G_0 = 2$, remaining parameters in both cases: $\Delta K = 8$, $\Delta G = 2$.
(c) $\Delta K = 3$, $\Delta G = 10$, $G_0 = 2$ [cf. Eq. \eqref{parametrization}]. In all panels, dots are obtained by integrating the full system \eqref{model} with $N = 10^4$ oscillators. All lines are obtained by numerically solving Eqs. \eqref{third_cond_1}-\eqref{r2}.
In order to unfold the hysteresis in (b), besides forward continuation $\left(\triangleright\right)$ as in the other cases, a backward
continuation $\left(\triangleleft\right)$ is performed. Use of colored regions as in Fig. \ref{Fig2}.}
\label{Fig3}
\end{figure*}
Finally, as outlined below, the intersection given by $G_{2,1}=0$ and \eqref{Kcrit3} coincides with the origin of bistability.
Note that the aforementioned conditions are exact. Figures \ref{Fig2}(a) and \ref{Fig2}(b) depict bifurcation diagrams in the planes spanned by the 
coupling mismatches $\left(\Delta K,\ \Delta G\right)$ and the average coupling strengths $\left(K_0,\ G_0\right)$, respectively. 
Solid lines are given by the critical conditions \eqref{Kcrit12} and \eqref{Kcrit3}. Dashed lines are obtained on the basis of the reduced system
\eqref{rtheta} with the help of $\texttt{MATCONT}$ \cite{DhGoKuz03}. We detect branch and limit points, since at all lines one eigenvalue vanishes, except at the ones given by \eqref{Kcrit12}, because those do not correspond to real bifurcations, but delineate two analogous partially synchronous states: zero-lag and $\pi$-states. We additionally test all these findings by numerically calculating the eigenvalues of the Jacobian of \eqref{rtheta} with the fixed points given by \eqref{third_cond_1}-\eqref{r2}. When the two lines given by \eqref{Kcrit12} and \eqref{Kcrit3} intersect, the boundaries \eqref{Kcrit12} cease to exist. We emphasize two distinct routes to TW states; TW1 is surrounded by $\pi$-states, TW2 by classical zero-lag sync states. Delimiting lines approach each other, see insets for enlarged areas. We further find bistability between incoherence and zero-lag or $\pi$-states, see panels (a) and (b) in Fig. \ref{Fig2}, respectively. The bistable areas are circumscribed by two lines that intersect at the points given by $\Delta G=\pm 2G_0$ and \eqref{Kcrit3}. The location of this intersection determines the type of bistability. Interestingly, traveling and non-traveling wave states can coexist in small parameter regions. We show this in Fig. \ref{Fig2}(c). In particular it is observed that traveling waves can coexist with complete incoherence, as well as with $\pi$-states and zero-lag partially synchronous states.

Figure \ref{Fig2} suggests certain conditions for the various collective states.
In order to observe $\pi$-states, mixed attractive-repulsive in-couplings have to be included. Traveling waves surrounded by $\pi$-states are possible, if one includes mixed in-couplings without mixed out-couplings. In contrast, traveling waves surrounded by zero-lag sync states exist, if there are mixed out-couplings and a non-zero mismatch without mixing in the in-couplings. Bistability between incoherence and zero-lag sync requires lack of mixed in-couplings, but the presence of mixed out-couplings. Finally, bistability between incoherence and $\pi$-states is possible by combining mixed in-couplings with vanishing mixing in the out-couplings. These conditions appear to supplement consistently the observations in Refs.
\cite{HoStr12,HoStr11PRL,*HoStr11}. In particular, we verified numerically that the traveling waves surrounded by zero-lag sync can also be observed in the setting studied in Ref. \cite{HoStr12}, if a small mismatch in the in-couplings is present, as suggested by our bifurcation diagram in Fig. \ref{Fig2}(a).

It is worth asking how crucial asymmetric interactions are for the discordant synchronization patterns discussed here. As it is easily seen from Eq. \eqref{parametrization}, the interactions are symmetric if in- and out-coupling strengths balance each other such that the equation $0=K_0\Delta G-G_0\Delta K$ holds. This condition can be projected onto straight cuts through the parameter space. The demarcations would not cross the traveling wave areas [see Fig. \ref{Fig2}(a) and \ref{Fig2}(b)]; instead they would divide the bifurcation diagrams into parameter regions that contain both routes to traveling waves, i.e. TW1 and TW2. In other words, asymmetric interactions are needed to get traveling waves. Interestingly however, those straight cuts would go through the $\pi$-state regimes. Thus, asymmetry in the interactions is not a necessary ingredient to observe $\pi$-states. This conclusion is not evident from the works presented in \cite{HoStr12,HoStr11PRL,*HoStr11}.

\section{Simulations}
Additional insights can be gained by numerically solving the three coupled equations \eqref{third_cond_1}-\eqref{r2} to get $r_{1,2}$ and $\delta$. The only subtlety is that one has to factor in the bifurcation values previously obtained in order to correctly choose between
\eqref{third_cond_1} and \eqref{third_cond_2}. Solutions are shown in Fig. \ref{Fig3}, and compared with the results from numerical simulations. 
For the latter, initial phases are randomly chosen from the uniform distribution $\left[-\pi,\pi\right]$. For each $K_0$ value a long-time average is taken over $t\in[2500,5000]$ with integration time step $dt=0.01$. Upper panels depict the order parameters, while in the lower panels the corresponding phase lags are shown. The colored regions match those in Fig. \ref{Fig2} and discriminate the different collective states. In Fig. \ref{Fig3}(b) one can see that at the transition from incoherence to $\pi$-state the suborder parameters abruptly jump from zero to high values in a hysteretic manner. For very long time averages it is expected that the hysteresis is washed out due to noise-induced jumps between the two stable steady states. We do not report this here. In Fig. \ref{Fig3}(c), around $K_0\approx -3$, the abrupt change in the phase lag and the non-vanishing order parameters signal extended stability of traveling waves, as discussed for Fig. \ref{Fig2}(c). Such bistable dynamics appears to be a promising topic for future studies. In general, the abrupt changes and the local minima in the order parameters as a function of the average in-coupling strength $K_0$, as presented in Fig. \ref{Fig3}, are of vital interest on their own, see e.g. Refs. \cite{pazo2005thermodynamic,*GogaGoArMo11} and \cite{OmelWo12}, respectively.
\begin{figure}[!tpb]
\begin{center}
\includegraphics[width=0.9\linewidth]{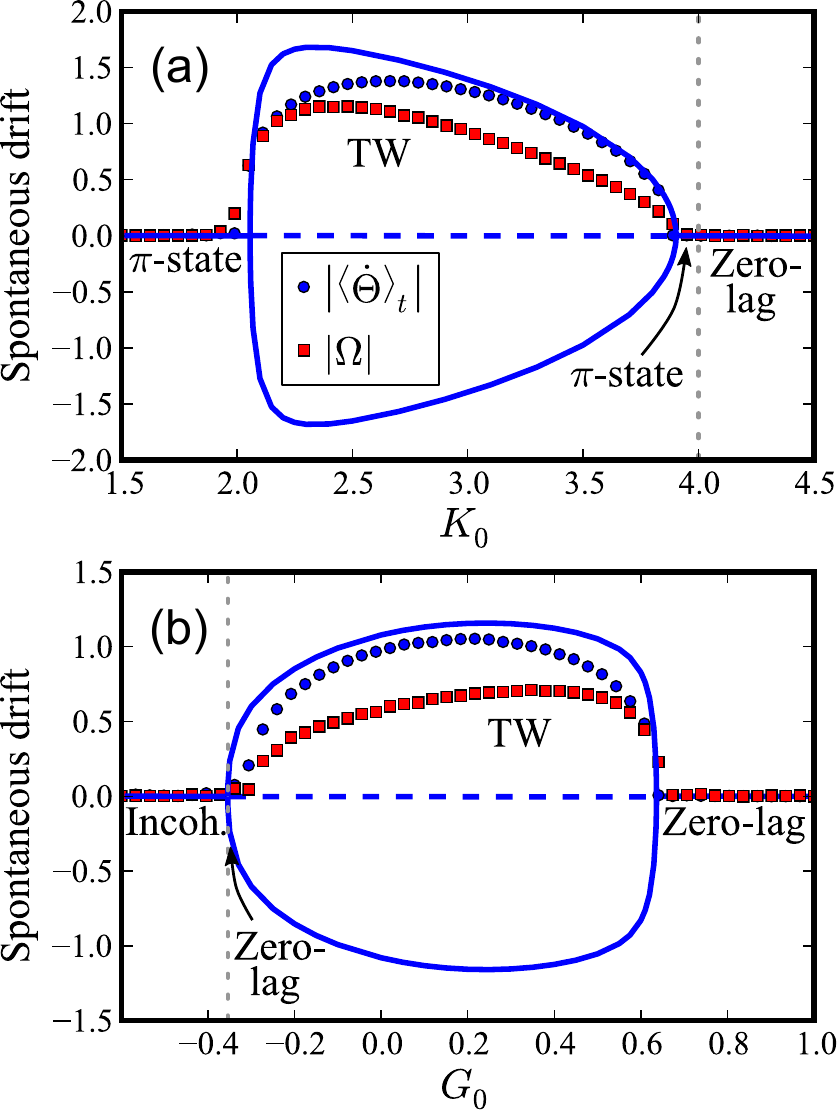}
\end{center}
\caption{(Color online) The spontaneous drift measured by common frequency $\left|\langle\dot\Theta\rangle_t\right|$, Eq. \eqref{common_freq} (theory [lines] vs. simulation [dots]) and wave speed $\left|\Omega\right|$, Eq. \eqref{wavespeed} (only simulation). (a) $G_0=2$, (b) $K_0=6$; remaining parameters: $\Delta K=8$, $\Delta G =2$ [cf. Eq. \eqref{parametrization}]. Simulation of $N = 10^4$ oscillators, averaging over $t\in[200,500]$, step size $dt = 0.01$.}
\label{Fig4}
\end{figure}

In Fig. \ref{Fig4} we compare the common frequency obtained from the reduced system, Eq. \eqref{common_freq}, with numerical simulations. It serves as an alternative measure to the wave speed \eqref{wavespeed}, which is calculated from the individual instantaneous frequencies that do not exist in the analytical treatment. Therefore for the wave speed no comparison with theory is being made. In Fig. \ref{Fig4} zero-lag synchronous and $\pi$-states become unstable in the regime of traveling waves. One can observe that both measures, the common frequency and the wave speed, highlight equally well the onset of traveling waves. As mentioned before, the waves emerge in frequency pairs, meaning that they can travel in both directions, depending on realization of random numbers.
\begin{figure*}[!tpb]
\begin{center}
\includegraphics[width=0.73\linewidth]{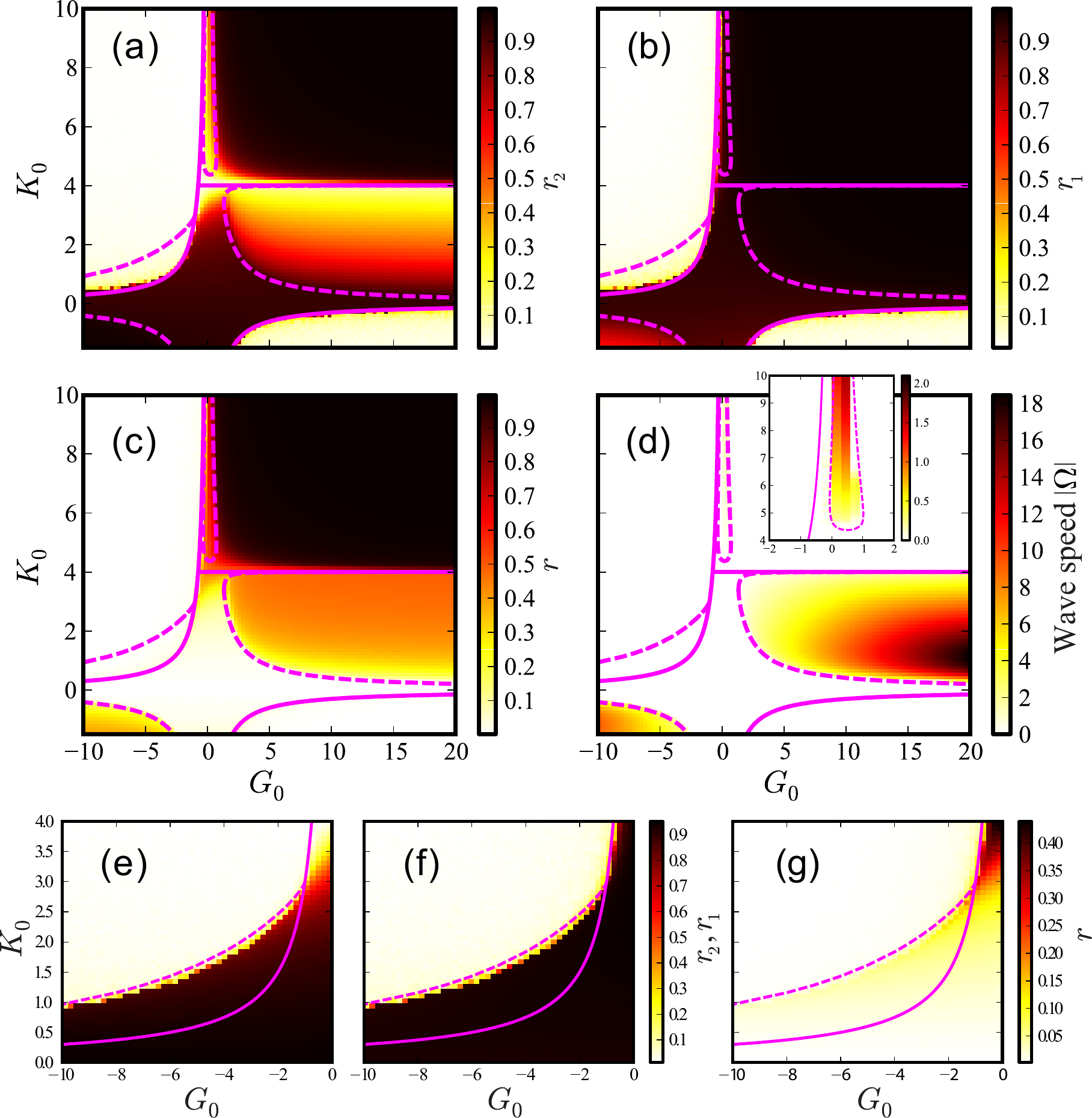}
\end{center}
\caption{(Color online) Simulations (colormaps) vs. theory (lines); panels (a) and (b) show local order parameters $r_2$ and $r_1$, while in (c) the global order $r$ is depicted. Panel (d) depicts the absolute value of the wave speed $\left|\Omega\right|$. Inset shows a zoom-in. Simulation performed with initial phases randomly chosen from uniform distribution $\left[-\pi,\pi\right]$. In panels (e)-(g) simulations are repeated with initial phases $\phi_i(0) = 1\ \forall i$. Remaining parameters are the same as in Fig. \ref{Fig2} (b).}
\label{Fig5}
\end{figure*}

In Fig. \ref{Fig5} results of numerical simulations are superimposed on smaller theoretical bifurcation diagrams. 
In Figs. \ref{Fig5}(a)-\ref{Fig5}(c) local and global order parameters are depicted, while in Fig. \ref{Fig5}(d) the wave speed is plotted. For each of the $100\times 100$ data points in the $K_0 \times G_0$ grid, the equations of motion \eqref{model} are integrated with $N=10^4$ oscillators and observables are then averaged over time, $t \in [100,500]$. Different initial conditions are chosen in Figs. \ref{Fig5}(e)-\ref{Fig5}(g) in order to find bistability in numerical simulations ($50\times50$ data points there). Specifically, in Figs. \ref{Fig5}(e)-\ref{Fig5}(g) the area circumscribed by dashed and solid lines shows $\pi$-states: the local order parameters $r_{1,2}$ attain large values, but due to the anti-phase alignment given by the phase lag $\pi$, the total order $r$ is small. In Figs. \ref{Fig5}(a)-\ref{Fig5}(c) this area is filled with incoherence. Hence, numerical simulations agree again very well with theoretical results. We remark that bistability between incoherence and zero-lag synchronous states, as predicted by the theory, can be analogously found by varying the initial conditions (not shown).

\section{Conclusion}
We explored the rich dynamics that emerge from asymmetric in- and out-coupling strengths among two mutually globally coupled oscillator populations. As an illustrative example we considered identical noisy Kuramoto phase oscillators with non-uniform and mixed attractive-repulsive interactions. We observed that the two populations can partially synchronize in-phase and with a constant phase lag to each other. We referred to the latter as ``discordant synchronization." The phase lags induced spontaneous drifts. As a result, traveling waves were formed in which the whole population oscillated with a different frequency than the individual units. However, in the state of maximal discordance, where the two partially synchronized populations are anti-aligned to each other, the spontaneous drifts disappeared. Correspondingly, we revealed two distinct routes to traveling waves, one through diametral two-cluster states, the other one through classical one-cluster states. Since the latter are ubiquitously investigated in the literature, we expect the second route to traveling waves to be more prevalent in general. Appropriate experimental setups made up of two constituents are conceivable, realized e.g. with laser systems \cite{Wunsche2005,*Zamora-Munt2010}, metronomes \cite{martens2013chimera} or chemical Belousov-Zhabotinsky oscillators \cite{tinsley2012chimera}.

With the help of a Gaussian approximation in the infinite system size limit we derived a three-dimensional system of coupled ODE's. This reduced system allowed a thorough bifurcation analysis and further analytical treatment, in excellent agreement with numerical simulations of a large but finite number of oscillators. We found that physically relevant singularities constitute a significant part of the bifurcation scenario. We further showed which collective states can coexist. Our results help to understand the emergence of discrepancies between individual and collective rhythms, as is observed e.g. in neuronal networks \cite{wilson1972excitatory,*brunel2000dynamics}. Specifically, if the connection strengths were capable to slowly vary in time, one could expect temporal patterns reminiscent of the high-frequency rhythmic events observed in hippocampal networks \cite{Gloveli2005,*Hofer2015}. Attractive (positive) and repulsive (negative) couplings are often associated with excitatory and inhibitory connections among neurons. This is reasonable, since positive couplings tend to increase synchrony, which is also the case for excitatory connections in the brain. In contrast, negative and inhibitory connections have in common that they tend to decrease synchrony \cite{cumin2007generalising}.
Exceptions however exist, see e.g. Ref. \cite{van1994inhibition}. The individual out-coupling strengths considered here are particularly suitable to emulate the role of excitatory or inhibitory neurons. This is pointed out in Ref. \cite{HoStr12} by referring to Dale's principle, according to which a neuron releases the same set of neurotransmitters at all its synapses \cite{eccles1976electrical}. Such a comparison would become even more applicable by including an excitation threshold into the system \cite{KurSchu95,*ZaNeFeSch03,SonnZaNeiLSG13,*SoPeRoKuLSG14,hong2014periodic}. Whether the combination of mixed attractive-repulsive interactions on the level of in- and out-coupling strengths is experimentally relevant, remains an interesting topic for the future.

Furthermore, future work should approach real networks of networks by considering multiple populations \cite{Kivela2014,*Boccaletti2014}, see the Appendix for first steps. It would also be interesting to extend the present framework towards inertia \cite{PhysRevE.90.062810,*komarov2014synchronization} and imposed phase shifts \cite{IaMcSt14,*vlasov2014synchronization,Maistrenko2014,*Burylko2014}. Finally, in small oscillator populations additional peculiarities can be expected \cite{Maistrenko2014,*Burylko2014}.\\

\acknowledgments
Thanks to S. Milster for comments on the manuscript.
B.S. and L.SG. acknowledge funding from the Bundesministerium f\"ur Bildung und Forschung (BMBF) (BCCN II A3, grant 31401211).
T.K.DM.P. acknowledges FAPESP (grant 2012/22160-7) and IRTG 1740. F.A.R. acknowledges CNPq (grant 305940/2010-4), 
FAPESP (grant 2013/26416-9) and IRTG 1740. J.K. acknowledges IRTG 1740 (DFG and FAPESP).

\appendix*
\section{Arbitrary number of populations}
\label{appendix}
Here we discuss the transition from incoherence to partial synchrony in an arbitrary number of interacting populations of arbitrary sizes. 
To this end, oscillators with the same pair of in- and out-coupling strengths are again grouped into one population. 
Then the nonlinear Fokker-Planck equation for the one-oscillator probability density $\rho(\phi,t|K,G)$ reads
\begin{equation}
\begin{aligned}
\frac{\partial\rho(\phi,t|K,G)}{\partial t}=&\ D\frac{\partial^2\rho(\phi,t|K,G)}{\partial\phi^2}\\
&-\frac{\partial}{\partial\phi}\bigl[\bigl\{KR(t)\sin\left[\Theta(t)-\phi(t)\right]\bigr.\bigr.\\
&\times\bigl.\bigl.\rho(\phi,t|K,G)\bigr\}\bigr]\ .
\label{Sfpe}
\end{aligned}
\end{equation}
The global mean-field amplitude $R(t)$ and phase $\Theta(t)$ are given by
 \begin{equation}
  R(t)\mathrm e^{i\Theta(t)}=\langle\langle r_{K',G'}(t)\ G'\ \mathrm e^{i\Theta_{K',G'}(t)}\rangle\rangle.
  \label{Sorder_fpe}
\end{equation}
The averages $\langle\langle\ldots\rangle\rangle\equiv\int dK'\int dG'\ldots P(K',G')$ take into account all in- and out-coupling strengths, $K$ and $G$, via the corresponding joint probability distribution $P(K,G)$. 
In Eq. \eqref{Sorder_fpe} the averages are further taken over the local mean-field variables,
$
r_{K,G}(t)\mathrm e^{i\Theta_{K,G}(t)}=\int_{0}^{2\pi} d\phi'\mathrm e^{i\phi'} \rho\left(\phi',t|K,G\right).
$
By a linear stability analysis of the incoherent state, $\rho(\phi,t|K,G)=1/(2\pi)\ \forall K,G,t$, one can exactly derive the critical noise
intensity for the synchronization transition:
\begin{equation}
D_c=\frac{\langle\langle K'G' \rangle\rangle}{2}.
\label{exact}
\end{equation}
This result follows from adapting the derivation \cite{Sak88,*StrMir91} to the present case.
Above $D_c$ the whole ensemble is incoherent, below $D_c$ the incoherent state loses stability and partial synchrony is observed.  
Now we give an alternative derivation on the basis of the Gaussian approximation. It recovers the exact condition \eqref{exact} and 
yields an instructive generalization of the equations in the main part of the manuscript. 
Inserting the Gaussian approximation for all phase distributions $\rho\left(\phi,t|K,G\right)$ into Eqs. \eqref{Sfpe}-\eqref{Sorder_fpe},
we obtain a set of coupled differential equations for the local mean-field amplitudes $r_{K,G}(t)$ and mean phases $\Theta_{K,G}(t)$:
\begin{equation}
\begin{aligned}
\dot{r}_{K,G}=&\ -r_{K,G}D + \frac{1-r_{K,G}^4}{2}K \\
&\ \ \times\big\langle\big\langle r_{K',G'}G'\cos\left(\Theta_{K',G'}-\Theta_{K,G}\right)\big\rangle\big\rangle,\\
\dot{\Theta}_{K,G}=&\ \frac{r_{K,G}^{-1}+r_{K,G}^3}{2}K \\
&\ \ \times\big\langle\big\langle r_{K',G'}G'\sin\left(\Theta_{K',G'}-\Theta_{K,G}\right)\big\rangle\big\rangle.
\end{aligned}
\label{dotrtheta_d_o_coupling}
\end{equation}
Let us consider small perturbations $\delta r_{K,G}(t)$ of the incoherent state, $r_{K,G}(t)=0\ \forall K,G,t$. The perturbations may give rise to
zero-lag synchronous or $\pi$-states, $\Theta_{K',G'}-\Theta_{K,G}=m\pi,\ m=0,1$. Accordingly, we can separate the network of networks into two groups, $1$ and $2$, which contain subpopulations with coupling strength pairs $(K,G)$ that lead to the same mean phases $\Theta_{K,G}$. At the same time, the mean phases of the subpopulations in the groups $1$ and $2$ differ by $\pi$. Linearizing around the perturbations $\delta r_{K,G}(t)$ in the two groups separately, we obtain from Eqs. \eqref{dotrtheta_d_o_coupling}:
\begin{equation}
\begin{aligned}
\left[\dot{\delta r}_{K,G}\right]_1=& -\left[\delta r_{K,G}\right]_1 D + \frac{1}{2}K_1 \\
&\ \times\bigl[\big\langle\big\langle \delta r_{K',G'} G'\big\rangle\big\rangle_1-\big\langle\big\langle \delta r_{K',G'} G'\big\rangle\big\rangle_{2}\bigr],\\
\left[\dot{\delta r}_{K,G}\right]_2=& -\left[\delta r_{K,G}\right]_2 D - \frac{1}{2}K_2 \\
&\ \times\bigl[\big\langle\big\langle \delta r_{K',G'} G'\big\rangle\big\rangle_1-\big\langle\big\langle \delta r_{K',G'} G'\big\rangle\big\rangle_{2}\bigr].
\end{aligned}
\label{perturbation}
\end{equation}
Here, the two separated networks of networks are labeled by the indices $1$ and $2$. According to Eq. \ref{Sorder_fpe}
the two perturbations can be put together as $\delta R(t)C=\big\langle\big\langle \delta r_{K',G'} G'\big\rangle\big\rangle_1-\big\langle\big\langle \delta r_{K',G'} G'\big\rangle\big\rangle_{2}$, where $C$ is some constant coming from an arbitrary global mean phase. As a result we obtain
\begin{equation}
\dot{\delta R}(t)=\left[-D+\frac{1}{2}\langle\langle K'G'\rangle\rangle\right]\delta R(t),
\label{total_perturbation}
\end{equation}
which leads to the critical condition \eqref{exact}.

\bibliography{bibliography}

\end{document}